\documentclass{article}
\usepackage[utf8]{inputenc}
\usepackage[a4paper, total={6.5in, 9.5in}]{geometry}
\usepackage{multicol}
\usepackage{tabularx}
\usepackage{graphicx}
\usepackage{float}
\usepackage{caption}
\usepackage{subcaption}
\usepackage{multirow}
\usepackage{textcomp}
\usepackage{amsmath}
\providecommand{\keywords}[1]{\textbf{\textit{Index terms---}} #1}
\usepackage[style=ieee]{biblatex}
\title{Evaluating Standard Feature Sets Towards Increased Generalisability and Explainability of ML-based Network Intrusion Detection}

\author{Mohanad Sarhan, Siamak Layeghy, Marius Portmann}
\date{\vspace{-5ex}}

\begin{document}

\maketitle

\begin{abstract}
Machine Learning (ML)-based network intrusion detection systems bring many benefits for enhancing the cybersecurity posture of an organisation. Many systems have been designed and developed in the research community, often achieving a close to perfect detection rate when evaluated using synthetic datasets. However, the high number of academic research has not often translated into practical deployments. There are several causes contributing towards the wide gap between research and production, such as the limited ability of comprehensive evaluation of ML models and lack of understanding of internal ML operations. This paper tightens the gap by evaluating the generalisability of a common feature set to different network environments and attack scenarios. Therefore, two feature sets (NetFlow and CICFlowMeter) have been evaluated in terms of detection accuracy across three key datasets, i.e., CSE-CIC-IDS2018, BoT-IoT, and ToN-IoT. The results show the superiority of the NetFlow feature set in enhancing the ML models detection accuracy of various network attacks. In addition, due to the complexity of the learning models, SHapley Additive exPlanations (SHAP), an explainable AI methodology, has been adopted to explain and interpret the classification decisions of ML models. The Shapley values of two common feature sets have been analysed across multiple datasets to determine the influence contributed by each feature towards the final ML prediction.

\end{abstract}

\keywords{
CICFlowMeter, Explainable, Machine Learning, NetFlow, Network Intrusion Detection System, SHaP
}

\section{Introduction}


Network Intrusion Detection Systems (NIDSs) aim to protect and preserve digital networks from cyber threats \cite{chaabouni2019network}. Traditional NIDSs aim to analyse incoming traffic signatures and match them to one of the known attack signatures \cite{garuba2008intrusion}. This method provides high detection accuracy of known attacks, however, it fails to detect unseen threats to computer networks, known as zero-day attacks \cite{garcia2009anomaly}. Therefore, researchers have applied emerging Machine Learning (ML) technologies to learn and detect the harmful patterns of network traffic to detect intrusions \cite{sinclair1999application}. A large number of academic research has been applied in this domain, many ML-based NIDSs have been developed, mostly achieving high detection accuracy when applied to certain datasets. However, the number of practical deployments of such systems in networks is vastly low \cite{sommer2010outside}. While several factors are contributing to the lack of translating academic research into operation, two main reasons seem to top the list; the lack of reliable comprehensive evaluation methodologies \cite{sarhan2021towards} and lack of understanding of internal ML-based NIDS operations \cite{amarasinghe2018toward}. In this paper, we investigate the possible solutions to both problems. 

First, the issue of unreliable ML-based NIDS evaluation methodology is addressed. Current benchmark NIDS datasets used in the testing stage of such systems are often generated with a unique set of network data features. Therefore, the capability of evaluating a learning model's detection accuracy using a targeted feature set across multiple datasets is limited. This questions the model's generalisability across other network environments and attack scenarios.  Most of the time, these systems and algorithms are evaluated via a single benchmark dataset, which can be attributed to the lack of availability of multiple benchmark datasets with a common feature set. A recent work \cite{sarhan2021towards} proposed a standard feature set, which enables the evaluation of ML-based NIDS across multiple datasets using a common NetFlow \cite{claise2004cisco} feature set. NetFlow was chosen due to its prevalence in the networking industry, however, it has not been validated as the best option to be used in NIDS datasets. The performance of ML models using the proposed feature set has never been compared to another set of features across multiple datasets. Therefore, in this paper, we examine the feature set designed by the CICFlowMeter \cite{lashkari2017characterization} to the NetFlow feature set proposed in \cite{sarhan2021towards} across three widely used datasets, i.e., CSE-CIC-IDS-2018, ToN-IoT and BoT-IoT. 

The comparison of the two feature sets (NetFlow and CICFLowMeter) has been conducted across three widely used NIDS datasets. Each dataset contains network data flows that have been collected over a different network environment. Moreover, various attack scenarios are available in each dataset. Hence, a fair and comprehensive comparison of the feature sets is provided in this paper. The NetFlow-based formats of the three datasets and the CICFLowMeter-based format of the CSE-CIC-IDS-2018 \cite{sharafaldin-habibi-lashkari-ghorbani-2018} dataset are available in the community. Therefore, in this paper two additional datasets have been generated in the CICFlowMeter format, named CIC-ToN-IoT and CIC-BoT-IoT, extracted using the ToN-IoT \cite{fesz-dm97-19} and BoT-IoT \cite{DBLP:journals/corr/abs-1811-00701} datasets, respectively. Both datasets have been labelled with binary- and multi-class categories and made publicly available at \cite{netflow_datasets_2020}. This will accommodate for a reliable evaluation of proposed ML models within the research community using the NetFlow and CICFlowMeter feature sets across multiple datasets. We strongly believe that such comprehensive evaluation methods will lead to reliable conclusions on the detection performance and generalisability of ML-based NIDS.

Moreover, we also examine the lack of understanding of internal ML operations \cite{arrieta2020explainable}. Currently, ML-based NIDSs are offered as a complex `black-box' \cite{mcgovern2019making} that achieves promising detection results on datasets. The complexity of ML models makes it hard to explain the reasoning behind the predictions made. Consequently, organisations are reluctant to trust ML decisions in a highly sensitive area such as intrusion detection \cite{amarasinghe2018toward}. Therefore, eXplainable Artificial Intelligence (XAI) methods are being applied in modern systems \cite{adadi2018peeking} to explain and interpret the decisions made by ML models. Explainable ML outputs can help maintain and troubleshoot the deployment of an ML-based NIDS by understanding and modifying the factors contributing to the model's decision. The classification results have been explained by calculating the Shapley value of each feature using the SHapley Additive exPlanations (SHAP) methodology \cite{lundberg2017unified}. SHAP measures the contribution and impact of each data feature towards the influence of the model's decision \cite{lundberg2017unified}. This will aid in the identification of key features utilised in the model's prediction and assist in the explanation of ML decisions.

Overall, this paper aims to address two key limitations of ML-based NIDS causing the wide gap between research and production. Two common feature sets based on NetFlow and CICFlowMeter have been evaluated across three widely used NIDS datasets. As part of the evaluation, two new datasets have been generated and made available at \cite{netflow_datasets_2020} to be used within the research community. Finally, an explanation of the ML classification results is conducted via the analysis of feature importance using the SHAP technique. In Section \ref{rw}, some of the key related works that focused on the generalisability and explainability of NIDS feature sets are discussed. In Section \ref{ds}, the importance of maintaining common feature sets is highlighted and the utilised datasets are introduced. The evaluation results are provided and analysed in Section \ref{results}. Section \ref{inter} describes the SHAP technique used to explain the importance and influence of network data features across the datasets. 

\section{Related Work}
\label{rw}

In this section, some of the key works that focused on explaining and evaluating the generalisability of feature sets of ML-based NIDS are discussed. In \cite{moustafa2015significant}, Moustafa et al. examined the attack detection performance of the UNSW-NB15 and KDD-99 datasets' feature sets. The experiment extracted the features of the UNSW-NB15 dataset from the KDD99 dataset. The Association Rule Mining (ARM) feature selection technique is applied to determine the most important features across the two datasets. The detection accuracy is measured using a Naive Bayes (NB) model and Expectation-Maximization (EM) clustering techniques. The results show that the original KDD-99 features are less efficient than the UNSW-NB15 features in terms of network attack detection. The NB and EM models achieved an accuracy of 62.02\% and 52.54\% using the original KDD-99 features and 78.06\% and 58.88\% using the UNSW-NB15 features, respectively. The authors did not replicate the KDD-99 features on the UNSW-NB15 dataset to reliably evaluate the two feature sets. 

In \cite{sarhan2020netflow}, Sarhan et al. highlighted the limitations of existing NIDS datasets, which is a lack of a common feature set, as the current feature sets are unique and completely different from each other. Therefore, the evaluation methods of the proposed ML-based NIDS are often unreliable. The lack of a common ground feature set prohibits the evaluation of the ML model's performance ability to generalise across multiple datasets. As a solution, the paper generated and published four datasets; NF-UNSW-NB15, NF-BoT-IoT, NF-ToN-IoT, and NF-CSE-CIC-IDS2018, sharing the same 12 NetFlow-based features. The datasets are generated by converting existing NIDS datasets to NetFlow format. NetFlow features are practical as they are relatively easier to extract from network traffic due to their presence in packet headers compared to complex features requiring deep packet inspection. As a use case, an Extra Tree classifier is reliably evaluated across the four datasets using the common feature set. The detection results indicate a reliable performance following a binary-class detection scenario. However, inferior results are achieved following a multiclass detection scenario.

In \cite{sarhan2021towards}, the authors extended the feature set proposed in \cite{sarhan2020netflow}, by extracting a total of 43 NetFlow-based features. The generated and labelled datasets named NF-UNSW-NB15-v2, NF-BoT-IoT-v2, NF-ToN-IoT-v2 and NF-CSE-CIC-IDS2018-v2 are generated. Their common feature set is proposed as a standard set to be used across future NIDS datasets. The authors argue the tremendous benefits of having a universal NetFlow-based standard feature set due to its wide usage in the networking industry, practical deployment, and scaling properties. An Extra Tree classifier is used to evaluate the proposed feature set and compare its detection accuracy with the basic NetFlow datasets generated in \cite{sarhan2020netflow} and the original feature set of the datasets. The results indicate that the proposed NetFlow feature set vastly outperforms the other feature sets in terms of attack detection accuracy across all datasets. The additional extracted NetFlow-based features demonstrated an extra amount of security events that enhance ML model detection rates of network attacks.

While the previous papers aimed to address the common feature set limitations faced by NIDS datasets, papers such as \cite{wang2020explainable} focused on explaining the internal operations of ML-based NIDS. The experimental methodology utilised the SHAP technique to interpret the decisions of the NIDS. To their knowledge, this is the first paper in the intrusion detection field to utilise the SHAP method in the understanding of the judgement and structure of the NIDS. The paper also explored the differences in interpretation between a binary- and multi-class classifier. The authors utilised a Deep Feed Forward (DFF) neural network with a ReLU activation function. The classifiers achieved an F1 score of 0.807 and 0.792 in binary- and multi-class experiments. However, the detection performance achieved by the model is lacking and the utilised single dataset, i.e., NSL-KDD dataset in the experiment, does not include or reflect the latest network threats and attacks which are commonly reported nowadays \cite{8672520}. The paper used a single attack class by selecting a random 100 samples of 'Neptune' attacks for the local explanation phase. The classifiers utilise different features to each other in their decision-making process. Finally, the top 20 important features for each attack type used in the global explanation are listed and comprehensively analysed via related works in the paper.

In \cite{mane2021explaining}, the paper designed a DFF and proposed an XAI framework to add transparency in ML operations. The DFF model is made up of 3 hidden layers performing a ReLU activation function and the output layer performing the softmax activation function. One aspect of the paper is the exploration of multiple XAI techniques, i.e., LIME, SHAP, Boolean Decision Rules via Column Generation (BRCG), and Contrastive Explanation Method (CEM). However, the paper utilised the NSL-KDD dataset to validate their methodology, which has been criticised in the field due to the lack of complex current network attack scenarios in the KDD-based datasets \cite{8672520}. The SHAP results deemed that a high value of the \textit{'same\_srv\_rate’} feature increases the probability of an attack prediction whereas a high value of the \textit{‘dst\_host\_serror\_rate’} feature increases the chances of a benign prediction. The authors applied the BRCG method to the dataset to extract model rules and achieved an accuracy of ~80\%. The LIME method generated a local explanation on specific data samples indicating which features contributed towards an attack or a benign prediction. Finally, the CEM showed how the prediction can be changed by altering feature values on a single benign data sample.

\section{Common NIDS Feature Set}
\label{ds}

NIDS datasets are made up of a number of network data features that reflect the information represented by datasets. These features are required to represent a sufficient amount of security events to aid the model's classification tasks. NIDS dataset features have a large impact on the final quality of the ML-based NIDS \cite{binbusayyis_vaiyapuri_2019}. In the deployment of ML-based NIDS in production, two key aspects require to be designed; a particular feature set for extraction and an ML model for feature analysis. The development of an optimal combination of a feature set and ML model has been an ongoing research issue. A large number of experiments targeting various feature selection and attack detection have been conducted \cite{ganapathy2013intelligent}. However, for a reliable evaluation of ML-based NIDS, multiple datasets are required to be utilised. As each dataset has been generated in a network environment where different attack types were conducted. Therefore, multiple datasets will aid in evaluating the generalisation ability of the ML-based NIDS detection capability in various attack scenarios and network environments.

The information represented by the datasets is determined by the choice of network features that make up the dataset. Currently, NIDS datasets have proprietary feature sets, which are unique in their design, and often completely different from each other. Dataset authors have applied their domain knowledge when selecting the network features. As a result, NIDS datasets are not similar in terms of network attributes represented to facilitate reliable experiments. It is important to note that it will be unfeasible to extract all of them, due to their large number, when the model is deployed over a practical network. The unique feature sets in datasets have restricted the evaluation of network data features across multiple datasets due to their absence in other datasets. It also prohibited the evaluation of an ML model across multiple datasets using a targeted feature set. Therefore, using a common feature set across multiple datasets is crucial in the design of the model, increasing the evaluation reliability and the chances of potential deployment.

\subsection{Current Datasets}

Common feature sets enable reliable experimental evaluation of ML models across various datasets and attack types. Currently, there are four NIDS datasets that share a common feature set \cite{sarhan2021toward}. The NetFlow-based features are proposed as a standard set to be used across future NIDS datasets. However, the performance of ML models using the NetFlow feature set has never been evaluated with another common set of features. Therefore, in this paper, the features designed by the CICFlowMeter tool \cite{lashkari2017characterization} are compared to the standard NetFlow feature set across three datasets. The key datasets (CSE-CIC-IDS2018, ToN-IoT, and BoT-IoT) discussed below have been widely used amongst the research community in the training and evaluation of ML-based NIDS. With the majority of the learning models achieving a reliable detection performance across the datasets. However, each dataset is made up of a unique, often completely different, set of network data features. Hence, the evaluation of such models has been conducted using different feature sets. This removes the conclusion of the model's generalisability to multiple datasets, each representing different network environments and attack types, using the proposed common feature set.

\begin{itemize}

    \item \textbf{CSE-CIC-IDS2018} \cite{sharafaldin-habibi-lashkari-ghorbani-2018}- A well-known NIDS dataset was released in 2018 in a project involving the Communications Security Establishment (CSE) \& Canadian Institute for Cybersecurity (CIC). The testbed used to emulate the network traffic was set up in an organisational network manner involving multiple departments. Attack types such as brute-force, bot, DoS, DDoS, infiltration, and web attacks were conducted from an external source. The dataset contains 75 features extracted using the CICFlowMeter-v3 tool \cite{lashkari2017characterization}. There are 16,232,943 data flows in total, where 13,484,708 (83.07\%) are benign and 2,748,235 (16.93\%) are attack samples.
    
    \item \textbf{ToN-IoT} \cite{fesz-dm97-19} - An IoT-based heterogeneous dataset released in 2019 that includes telemetry data of IoT services, network traffic of IoT networks, and operating system logs. The data was generated using an industrial network testbed. The dataset contains several attack scenarios such as backdoor, DoS, Distributed DoS (DDoS), injection, Man In The Middle (MITM), password, ransomware, scanning and Cross-Site Scripting (XSS). Bro-IDS tool, now called Zeek, was utilised to extract 44 network traffic features. 
    
    \item \textbf{BoT-IoT} \cite{DBLP:journals/corr/abs-1811-00701}- The Cyber Range Lab of ACCS designed an IoT based network environment that consists of normal and botnet traffic. The non-IoT and IoT traffic were generated using the Ostinato and Node-red tools, respectively. A total of 69.3GB of pcap files were captured and the Argus tool was used to extract the dataset's original 42 features. The dataset contains 477 (0.01\%) benign flows and 3,668,045 (99.99\%) attack ones, that is, 3,668,522 flows in total.
    
    \item \textbf{NF-CSE-CIC-IDS2018-v2} \cite{sarhan2021towards} - The CSE-CIC-IDS2018 dataset \cite{sharafaldin-habibi-lashkari-ghorbani-2018} have been converted into 43 NetFlow-based features using nProbe \cite{Ntopng2017} to generate NF-CSE-CIC-IDS2018-v2. There are 18,893,708 total number of flows where  2,258,141 (11.95\%) are attack samples and 16,635,567 (88.05\%) are benign ones. There are six attack categories such as bruteforce, bot, DoS, DDoS, infilteration, and web attacks.

    \item \textbf{NF-ToN-IoT-v2} \cite{sarhan2021towards} - An IoT dataset generated based on 43 NetFlow features released in 2021. The features are extracted using nProbe \cite{Ntopng2017} from the pcaps of the original parent (ToN-IoT) dataset \cite{fesz-dm97-19}, generated at the Cyber Range Lab by the Australian Centre for Cyber Security (ACCS). The total number of attack data flows 10,841,027 (63.99\%) and 6,099,469 (36.01\%) are benign dataflows, adding up to a total of 16,940,496 samples. There are nine attack categories known as backdoor, DoS, DDoS, injection, MITM, password, ransomware, scanning, and XSS.

    \item \textbf{NF-BoT-IoT-v2} \cite{sarhan2021towards} - A newly generated IoT dataset based on 43 NetFlow features was released in 2021. The features are extracted using nProbe \cite{Ntopng2017} from the pcaps of the original dataset, known as BoT-IoT, generated at the Cyber Lab of the ACCS \cite{DBLP:journals/corr/abs-1811-00701}. It contains 37,763,497 labelled network data flows, where the majority are attack samples; 37,628,460 (99.64\%) and, 135,037 (0.36\%) are benign. There are four attack categories in the dataset, i.e., DDoS, DoS, reconnaissance, and theft.

\end{itemize}

\subsection{Generated Datasets}

\begin{figure}[h!]
  \centering
  \includegraphics[width=12cm, height=2.5cm]{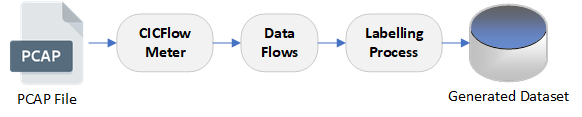}  
  \caption{Dataset Generation Process}
  \label{arc}
\end{figure}

In order to compare the performance of the NetFlow-based feature set to the CICFLowMeter-based feature set, three NIDS datasets are required to be available in both formats. The CSE-CIC-IDS2018 dataset is available natively in the CICFlowMeter format and NetFlow format as NF-CSE-CIC-IDS2018-v2. However, the ToN-IoT and BoT-IoT are only available in a NetFlow format as  NF-ToN-IoT-v2 and NF-BoT-IoT-v2, respectively. Therefore, in this paper, the CICFlowMeter tool \cite{lashkari2017characterization} has been utilised to convert the ToN-IoT and BoT-IoT datasets into the CICFLowMeter format. The packet capture (pcap) files have been fed into the CICFlowMeter tool that extracts the required feature set and generated data flows. The flows have been labelled in a binary- and multi-class manner, using ground truth events to allow for supervised learning methods. The source/destination IPs and ports and protocol features have been utilised to locate each data sample in the ground truth events and identify the respective attack type. Figure \ref{arc} presents the overall process used to generate the new datasets. The generated datasets using the ToN-IoT and BoT-IoT datasets have been named CIC-ToN-IoT and CIC-BoT-IoT, respectively, and made publicly available for research purposes at \cite{netflow_datasets_2020}.

\begin{itemize}

    \item \textbf{CIC-ToN-IoT}- A dataset generated as part of this paper, where the feature set of the CICFlowMeter was extracted from the pcap files of the ToN-IoT dataset \cite{fesz-dm97-19}. The CICFlowMeter-v4 tool \cite{lashkari2017characterization} was utilised to extract 83 features. There are 5,351,760 data samples where 2,836,524 (53.00\%) are attacks and 2,515,236 (47.00\%) are benign samples. This dataset has been generated as a part of this paper and made available at \cite{netflow_datasets_2020}.
    
    \item \textbf{CIC-BoT-IoT}- The CICFlowMeter-v4 \cite{lashkari2017characterization} was used to extract 83 features from the BoT-IoT dataset \cite{DBLP:journals/corr/abs-1811-00701} pcap files. The dataset contains 13,428,602 records in total, containing 13,339,356 (99.34\%) attack samples and 89,246 (0.66\%) benign samples. The attack samples are made up of four attack scenarios inherited from the parent dataset, i.e., DDoS, DoS, reconnaissance, and theft. This dataset has been published as a part of this paper at \cite{netflow_datasets_2020}.
\end{itemize}

The generated datasets share the same feature set as the CSE-CIC-IDS2018 dataset, proposed by the CICFLowMeter. That provides a reliable comparison of the feature set across three different datasets. It also enables the evaluation of ML models across the datasets using a common feature set. The chosen datasets in this paper; NF-CSE-CIC-IDS2018-v2, NF-ToN-IoT-v2, and NF-BoT-IoT-v2 in a NetFlow-based format and CSE-CIC-IDS2018, CIC-ToN-IoT and CIC-BoT-IoT in a CICFlowMeter format will allow for an evaluation of ML-based NIDS on two common feature sets across multiple network environments. It will also comprehensively evaluate both feature sets and compare the performances of enabling ML models for the detection of a large variety of network scenarios. We believe that having two common feature sets across six different datasets will significantly assist the research community in the evaluation of the proposed system. Consequently, tightening the gap between academic research and practical deployments.

\section{Evaluation}
\label{results}
In this section, two common network data feature sets are being evaluated across six NIDS datasets. This includes three datasets in the CICFlowMeter format (CSE-CIC-IDS2018, CIC-BoT-IoT, and CIC-ToN-IoT) and their three respective datasets in the NetFlow format (NF-CSE-CIC-IDS2018-v2, NF-BoT-IoT-v2, and NF-ToN-IoT-v2). To the best of our knowledge, this is the first-ever evaluation of ML-based NIDS using two common feature sets across recent multiple NIDS datasets. The large variety of network traffic in each dataset includes different attack scenarios and benign applications that will extensively and comprehensively evaluate the combination of feature sets and ML models, which generally form the key components of an ML-based NIDS.

\subsection{Methodology}
During the experiments, Deep Feed Forward (DFF) and Random Forest (RF) classifiers are utilised to classify the network data flows present in the datasets. The DFF structure consists of an input layer and three hidden layers with 10 nodes each performing the ReLU activation function. The output layer is a single Sigmoidal node and the Adam optimisation algorithm is used. The RF model consists of 50 randomised decision trees, each performing the Gini function to measure the quality of a split. To avoid learning bias towards the attack and victim devices, the source/destination IPs and ports are dropped. In addition, the timestamp and flow ID features are removed as they are unique to each data sample. The min-max scaler is applied to normalise all values between zero and one. This is necessary to avoid increased attention to larger values. Several binary classification metrics are collected, including accuracy,  \textit{F1 Score}, \textit{Detection Rate (DR)}, \textit{False Alarm Rate (FAR)}, \textit{Area Under the Curve (AUC)}, and the \textit{Prediction Time} required to predict a single data sample in microseconds (\textmu s).  The datasets have been split into 70\%-30\% for training and testing purposes. For a fair evaluation, five cross-validation splits are conducted and the mean results are measured.



\subsection{Results}

\begin{table}[ht] \scriptsize
\centering
\caption{RF classification results}
\label{tab:my-table}
\resizebox{\textwidth}{!}{%
\begin{tabular}{l|r|r|r|r|r|r|}
\cline{2-7}
 &
  \multicolumn{1}{l|}{\textbf{Accuracy}} &
  \multicolumn{1}{l|}{\textbf{F1 Score}} &
  \multicolumn{1}{l|}{\textbf{DR}} &
  \multicolumn{1}{l|}{\textbf{FAR}} &
  \multicolumn{1}{l|}{\textbf{AUC}} &
  \multicolumn{1}{l|}{\textbf{Prediction Time}} \\ \hline

\multicolumn{1}{|l|}{\textbf{NF-CSE-CIC-IDS2018-v2}} & 99.47\% & 0.98 & 96.82\% & 0.17\% & 0.9833 & 20.98\textmu s \\
\multicolumn{1}{|l|}{\textbf{CSE-CIC-IDS2018}}       & 98.01\% & 0.93 & 94.75\% & 1.42\% & 0.9667 & 22.39\textmu s \\ \hline
\multicolumn{1}{|l|}{\textbf{NF-ToN-IoT-v2}}         & 99.66\% & 1.00    & 99.80\% & 0.58\% & 0.9961 & 7.57\textmu s  \\
\multicolumn{1}{|l|}{\textbf{CIC-ToN-IoT}}           & 99.33\% & 0.99 & 99.80\% & 1.22\% & 0.9929 & 10.14\textmu s \\ \hline
\multicolumn{1}{|l|}{\textbf{NF-BoT-IoT-v2}}         & 100.00\%   & 1.00    & 100.00\%   & 0.25\% & 0.9988 & 3.60\textmu s  \\
\multicolumn{1}{|l|}{\textbf{CIC-BoT-IoT}}           & 98.24\% & 0.99 & 98.24\% & 1.53\% & 0.9836 & 7.07\textmu s  \\ \hline
\end{tabular}%
}
\end{table}

The DFF and RF classifiers are utilised to classify the dataset samples into attack and benign categories. Tables \ref{tab:my-table} and \ref{tab:my-table2} list the attack detection results of the six datasets using the RF and DFF classifiers respectively. Both classifiers have achieved a higher detection accuracy on the NF-CSE-CIC-IDS2018-v2 dataset compared to the CSE-CIC-IDS2018 dataset by increasing the DR and lowering the FAR. Resulting in an increase of the F1 score from 0.93 to 0.98 and 0.90 to 0.97 using the RF and DFF classifiers respectively.  These affirm the results published in \cite{sarhan2021towards}. The RF classifier achieved very similar detection results on the ToN-IoT and BoT-IoT datasets in both their NetFlow and CICFlowMeter feature sets. The F1 score increases from 0.99 to 1.00 in both datasets. The FAR drops from 1.22\% in CIC-ToN-IoT to 0.58\% in NF-ToN-IoT-v2 and from 1.53\% in CIC-BoT-IoT to 0.25\% in NF-BoT-IoT-v2. It is also noted that the NetFlow features require a significantly lower prediction time compared to the CICFlowMeter features in both datasets. The DFF model achieved a significantly increased DR in the NF-ToN-IoT-v2 dataset of 95.37\% compared to 92.29\% in the CIC-ToN-IoT dataset, which resulted in an F1 Score increase from 0.94 to 0.96. This is also the scenario in the NetFlow features of the NF-BoT-IoT-v2 dataset, in which a high DR of 99.54\% is achieved compared to CIC-BoT-IoT of only 95.99\% DR.

\begin{table}[ht]\scriptsize
\centering
\caption{DFF classification results}
\label{tab:my-table2}
\resizebox{\textwidth}{!}{%
\begin{tabular}{l|r|r|r|r|r|r|}
\cline{2-7}
 &
  \multicolumn{1}{l|}{\textbf{Accuracy}} &
  \multicolumn{1}{l|}{\textbf{F1 Score}} &
  \multicolumn{1}{l|}{\textbf{DR}} &
  \multicolumn{1}{l|}{\textbf{FAR}} &
  \multicolumn{1}{l|}{\textbf{AUC}} &
  \multicolumn{1}{l|}{\textbf{Prediction Time}} \\ \hline

\multicolumn{1}{|l|}{\textbf{NF-CSE-CIC-IDS2018-v2}} & 99.24\%             & 0.97              & 94.67\%       & 0.14\%         & 0.9831       & 8.23\textmu s                     \\
\multicolumn{1}{|l|}{\textbf{CSE-CIC-IDS2018}} & 97.05\% & 0.90  & 85.71\% & 0.96\% & 0.9502 & 8.31\textmu s \\ \hline
\multicolumn{1}{|l|}{\textbf{NF-ToN-IoT-v2}}   & 94.74\% & 0.96 & 95.27\% & 6.08\% & 0.9843 & 8.12\textmu s \\
\multicolumn{1}{|l|}{\textbf{CIC-ToN-IoT}}     & 93.80\%  & 0.94 & 92.29\% & 4.49\% & 0.9782 & 7.26\textmu s \\ \hline
\multicolumn{1}{|l|}{\textbf{NF-BoT-IoT-v2}}   & 99.54\% & 1.00    & 99.54\% & 0.20\%  & 0.9996 & 8.37\textmu s \\
\multicolumn{1}{|l|}{\textbf{CIC-BoT-IoT}}     & 96.01\% & 0.98 & 95.99\% & 1.20\%  & 0.9907 & 9.88\textmu s \\ \hline
\end{tabular}%
}
\end{table}

\begin{figure*}[ht]
\begin{subfigure}{.5\textwidth}
  \centering
  \includegraphics[width=9cm, height=5cm]{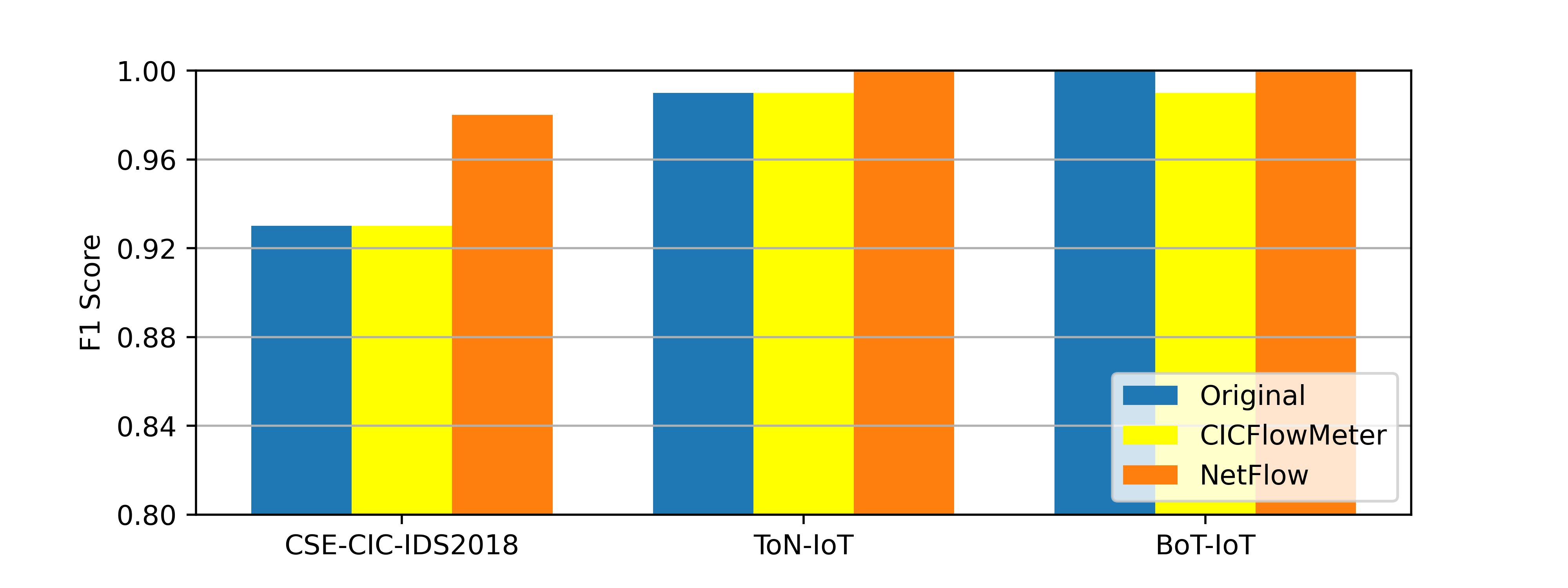}  
  \caption{RF}
  \label{uu}
\end{subfigure}
\hfill
\begin{subfigure}{.5\textwidth}
  \centering
  \includegraphics[width=9cm, height=5cm]{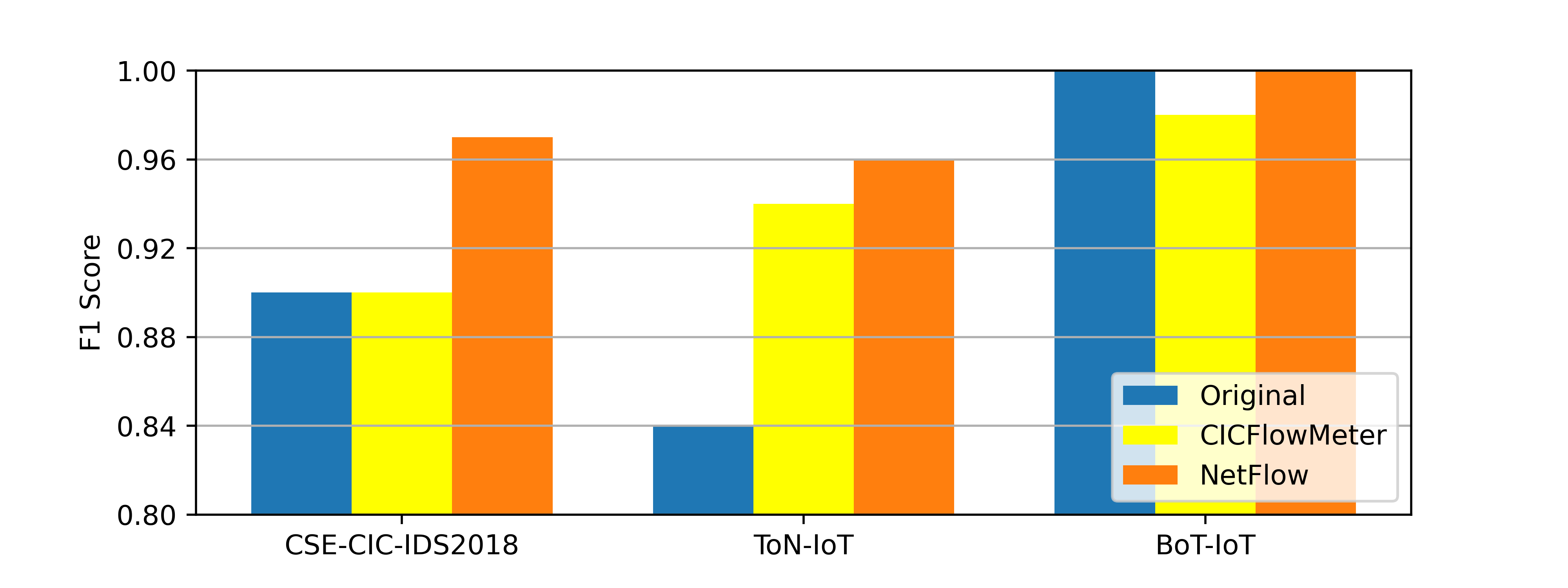}  
  \caption{DFF}
  \label{u}
\end{subfigure}
\caption{Classification performance of two feature sets across three NIDS datasets}
\label{tt}
\end{figure*}

The lower prediction time required by the NetFlow features compared to the CICFlowMeter features can be explained by the smaller total number of features that make up the dataset. The higher DR of attacks and lower FAR indicates that the NetFlow-based features contain more amount or better quality of security events that aid the ML models efficient network intrusion detection. Overall, the constantly higher achieved detection accuracies demonstrate that the proposed NetFlow feature set in \cite{sarhan2021towards} is in a better capacity to assist ML models in identifying the attacks present in all three datasets. Figure \ref{tt} visually presents the benefits of using the NetFlow feature set in attack detection scenarios compared to the CICFlowMeter feature set. The results are grouped by Figures \ref{uu} and \ref{u} based on the DFF and RF classifiers respectively. The proprietary feature set of the ToN-IoT and BoT-IoT datasets has been added for comparison purposes, comparing both datasets in their original, NetFlow and CICFlowMeter formats. The CSE-CIC-IDS2018 dataset has been released in the CICFlowMeter format as its proprietary format. Therefore, the original and CICFlowMeter formats are identical. The F1 score is plotted on the y-axis and the datasets on the y-axis, the figures show that the NetFlow features constantly achieve a higher F1 score across the three datasets compared to the CICFlowMeter and original feature sets.

The continuous improvement of the ML models' detection accuracy, when trained and evaluated using the NetFlow feature set over the CICFlowMeter features, demonstrates the advantages of standardising it in the NIDS community. The NetFlow feature set enables two ML models; DFF and RF, following deep and shallow learning structures, to achieve a reliable performance over three datasets designed over different network environments and include a large number of different attack types. The CICFlowMeter results are slightly less superior in the BoT-IoT and ToN-IoT datasets and less efficient to a certain extent in the detection of attacks in the CSE-CIC-IDS2018 dataset. Moreover, another advantage of the NetFlow feature set is it contains a lower number of features of 43 compared to the CICFlowMeter of 83 features. A lower number of features will aid in an enhanced operation of extraction, analysis, and storage of features over an operational network. Finally, the NetFlow features are naturally present in network packet headers that do not require additional tasks to collect, unlike the CICFlowMeter features that include statistically measured features based on sum, maximum, minimum, standard deviation and average calculations, which might be unfeasible to generate in a live high-speed network.

\section{Explainable ML-based NIDS}
\label{inter}
  
In this section, the classification results achieved above will be analysed and explained to identify the key features that contributed to the models' final predictions. While a great deal of effort has been put into the design of ML-based NIDSs to achieve great detection accuracy, the trust in their operation of securing computer networks has always been in question \cite{amarasinghe2018toward}. As a result, a very limited number of commercial NIDSs are equipped with ML capabilities. It is believed this is due to the nature of such tools where the cost of errors are significantly higher compared to other domains \cite{sommer2010outside}, which means the operation of NIDS requires to be accurate at all times to avoid significant disruption. ML is often known to be a 'black box' technology where there is no clear understanding of what patterns are learned or why predictions are made. In the context of ML-based NIDS, the way models train, predict, and classify network traffic into an attack category is often not transparent. Therefore, organisations are reluctant to implement an ML-based tool  \cite{amarasinghe2018toward}. There are multiple reasons such as data source, data features, class imbalance, and inaccurate datasets that can have a vast impact on the predictions made by ML models \cite{amarasinghe2018toward}. This makes it critical to understand and gain insights into the ML internal operations and decisions made.

The above motivation led to the generation of an ongoing research area of eXplainable Artificial Intelligence (XAI), where the aim is to analyse and explain the internal operations of ML models. As such, implementing the technologies of XAI into the ML-based NIDS is important to tighten the gap between the extensive academic research conducted and the number of operational deployments by increasing trust in ML. One way to interpret the detection results of an ML model is to determine which network data features contribute to the decision of the classifier. Determining which features of the dataset were used by the classifier to differentiate between a benign and an attack sample is critical for many reasons. Firstly, it helps in identifying which features contain security events that should be used in the design of the model or a better feature set. On the other hand, invaluable features that contain a limited number of security events and should be omitted from datasets are also located. Moreover, the classification decision of the model will be justified based on the values of these features. This helps in troubleshooting the errors caused by the wrong predictions of the model, as it allows security experts to analyse the values of the influencing features that result in a miss-classification. Further tuning of the model parameters and utilised features can be conducted after such information.

\subsection{Shapley Analysis}
\label{SHAP}
The interpretation of the model's decisions through the calculation of each feature contribution will help in uncovering the 'black box' of ML. In this paper, the \textit{Shapley values} are used to explain the importance of network data features in the detection of network attacks. The Shapley value was invented by Lloyd Shapley in 1953 \cite{shapley201617}, which is the method of assigning players' payouts based on their contribution to the total game's payout. The theory behind the Shapley value has been adopted in ML, where the `game' represents the prediction task of a single sample in a dataset. The `payout' is the actual prediction for one sample minus the average predictions of all samples. The `players' are the feature values of the samples that collaborate to obtain the `payout'. Overall, the Shapley value is the weighted average of the respective contribution of a feature value. The Shapley value ($\phi_{j}$) of feature $j$ is defined via Equation \ref{eq1} \cite{kreps1990game}, where $S$ is a subset of features (values) used by the model, $X=\{x_j/\ $j$ \in [1,...,$p$]\}$ represents the vector of feature values of the dataset samples in which  $p$ is the number of total features and $x_{j}$ is a feature value, and $val_x(S)$ is the final prediction for feature values in a test set $S$. 

\begin{equation}
\phi_{j}(v a l)=\sum_{S \subseteq\left\{x_{1}, \ldots, x_{p}\right\} \backslash\left\{x_{j}\right\}} \frac{|S| !(p-|S|-1) !}{p !}\left(\operatorname{val}\left(S \cup\left\{x_{j}\right\}\right)-\operatorname{val}(S)\right)
\label{eq1}
\end{equation}

SHapley Additive exPlanations (SHAP) is a common XAI technique developed by Lundberg and Lee \cite{lundberg2017unified}. It is based on an additive feature importance method of calculating the Shapley values, known as \textit{KernelSHAP} and \textit{TreeSHAP}. Compared to other XAI methods, SHAP has a strong theoretical foundation and can be used to explain the output of any ML model.  SHAP presents new methods that have shown enhanced performance compared to alternative techniques. KernelSHAP is a kernel-based calculation approach for Shapley values, inspired by local surrogate models. TreeSHAP is used to explain tree-based ML models such as decision trees, random forest, and extra trees by leveraging their internal `tree' structure to speed up the explanation process. When implemented, SHAP explains the ML prediction of each data sample $x$ by calculating the importance of each feature based on its contribution to the ML prediction process. Equation \ref{eq2} defines the explanation as specified by SHAP, 
\begin{equation}
g\left(z^{\prime}\right)=\phi_{0}+\sum_{j=1}^{M} \phi_{j} z_{j}^{\prime}
\label{eq2}
\end{equation}
where $g$ is the explanation model, $z^{\prime} \in\{0,1\}^{M}$ is the coalition vector of features used, $M$ represents the maximum coalition size, and $\phi_j \in R$ is the Shapley value of feature $j$. When using SHAP to determine feature importance, the features with larger Shapley values are more important. SHAP calculates the average of importance per feature across the dataset using Shapley values. The Shapley value defines the amount of contribution of a feature value per data sample. The mean Shapley value is the average of the Shapley values across all test samples. A higher mean Shapley value indicates a stronger influence of the feature value over the final prediction of the model. Hence, a more important feature to analyse and investigate is indicated by a larger Shapley value. 

\subsection{Results}

While evaluating the RF classifier using the CSE-CIC-IDS2018 dataset, network data features containing security events relating to the forward direction of the flow make up the top four features influencing the model's decision. In particular, the \textit{'Fwd Seg Size Min'} feature has almost double the influence of any other feature. For the DFF classifier, the top two features present the number of forward and backward packets per second, followed by the \textit{'Fwd Seg Size Min'} feature which is the most influencing feature in the RF classifier's decision. The Shapley analysis of the NF-CSE-CIC-IDS2018-v2 dataset using the RF and DFF classifiers has 13 common features in the top 20 influencing features for both classifiers. This indicates that they contain key security events that can be utilised in the detection of attacks present in the NF-CSE-CIC-IDS2018-v2 dataset. The common features mainly include Transmission Control Protocol (TCP)- and Time To Live (TTL)-based features. The \textit{'TCP\_WIN\_MAX\_OUT'} is the first and second most influencing feature in the DFF and RF classifiers respectively, where it presents the maximum TCP window from the destination to the source host. In the analysis of the CIC-ToN-IoT dataset, the RF classifier has determined that the \textit{'Idle Mean, Min, and Max'} as the key features, influencing more than 50\% of the model's final predictions. The features present the average, minimum, and maximum time the flow was idle before becoming active. However, the DFF failed to fully utilise neither the idle-based features in the attack detection stage. Where there are only two idle-based features in the top 20 list. This can explain the lower performance of the DFF classifier compared to the RF classifier. Further analysis of the idle-based features is required.

\begin{figure*}[h!]
\begin{subfigure}{.5\textwidth}
  \centering
  \includegraphics[width=7cm, height=5.5cm]{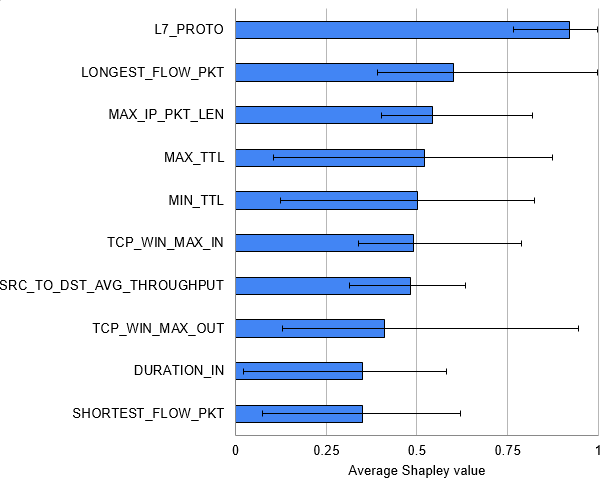}  
  \caption{RF}
  \label{7}
\end{subfigure}
\hfill
\begin{subfigure}{.5\textwidth}
  \centering
  \includegraphics[width=7cm, height=5.5cm]{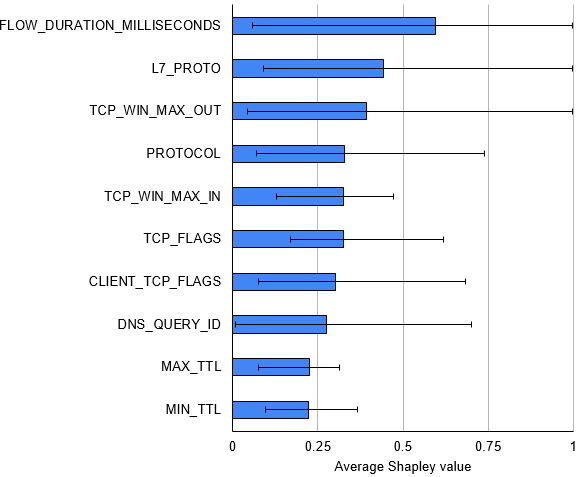}  
  \caption{DFF}
  \label{77}
\end{subfigure}
\caption{Average SHAP values of three NetFlow-based datasets}
\label{nf}
\end{figure*}

\begin{figure*}[h!]
\begin{subfigure}{.5\textwidth}
  \centering
  \includegraphics[width=7cm, height=5.5cm]{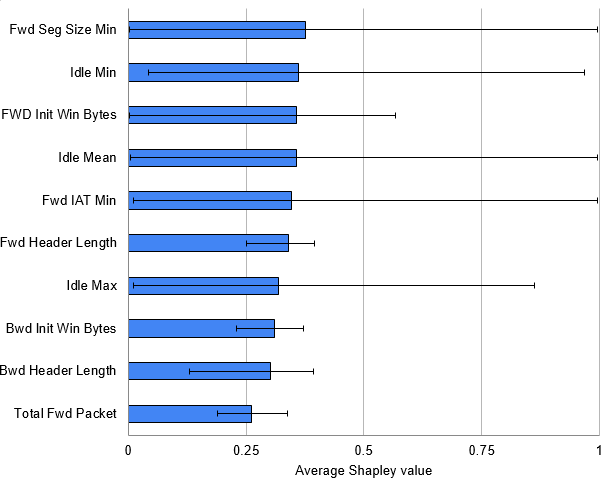}  
  \caption{RF}
  \label{8}
\end{subfigure}
\hfill
\begin{subfigure}{.5\textwidth}
  \centering
  \includegraphics[width=7cm, height=5.5cm]{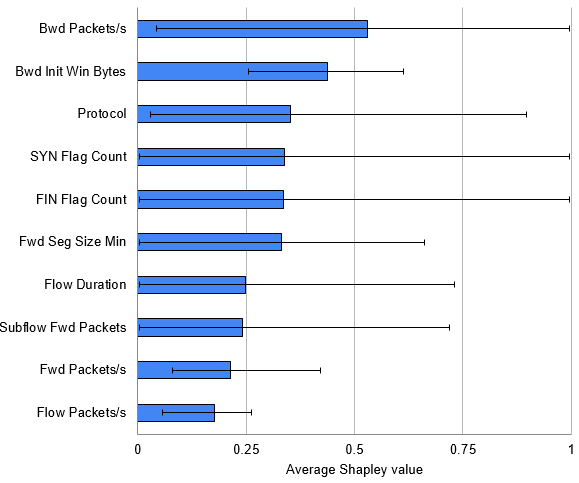}  
  \caption{DFF}
  \label{88}
\end{subfigure}
\caption{Average SHAP values of three CICFlowMeter-based datasets}
\label{cic}
\end{figure*}

In the NF-ToN-IoT-v2 and NF-BoT-IoT-v2 datasets, there are 17 and 14 common features out of the top 20 features influencing the RF and DFF classifiers, respectively, this indicates that they withhold key security events to aid the ML model detection performance. In the CIC-BoT-IoT dataset, the forward directional-based security events are presented in four out of the top five features that impact the RF classifier decisions. The Information Access Technology (IAT)-based features account for nine out of the top 20 features that aid the RF classifier to detect attacks present in the CIC-BoT-IoT dataset. In particular, the \textit{'Fwd IAT Min and Mean'} features are the top two, representing the minimum and average time between two packets sent in the forward direction. The DFF classifier likewise utilised the IAT-based features that made up eight features out of the top 20. Moreover, the average SHAP values of  NetFlow and CICFlowMeter feature sets are calculated across the three NIDS datasets. The top 10 features of the NetFlow and CICFlowMeter feature sets are displayed in Figures \ref{nf} and \ref{cic} respectively. The features (y-axis) are ranked based on their mean Shapley values (x-axis) across the whole test data samples used to evaluate the ML models. Each figure presents the mean Shapley values as determined by the KernelSHAP and TreeSHAP methods for the DFF and RF models, respectively. For efficient comparison, the mean Shapley values have been normalised to a scale from 0 to 1. The error bars are set based on the maximum and minimum SHAP values across the three datasets. The large margin of error across both feature sets, particularly in the CICFlowMeter format, illustrates that there is no optimal choice of the most influencing features across the utilised datasets. However, there are some similarities in Figure \ref{nf}, where the \textit{L7\_PROTO}, TCP- and TTL-based features tend to be the most influencing features on both ML classifiers across the NetFlow-based features. It is also noticeable that in the case of the NetFlow feature set, there are 5 common features in the top features of deep and shallow learning methods, while for the FlowMeter format it is only 2 features.

Overall, a global interpretation has been conducted by analysing the Shapley values of six NIDS datasets. The datasets sharing two common feature sets (CICFlowMeter and NetFlow) have been utilised to analyse the impact of each feature across multiple datasets. The KernelSHAP and TreeSHAP methods have been utilised for the DFF and RF models, respectively. The results identify certain key features causing a major impact on the model detection. However, some features have been vastly insignificant in the classification's process. This relates to the amount and quality of security events that these features present relating to the detection of attack types in the dataset. Therefore, the amount of each feature's contribution is different based on the datasets. This could be explained by the different attack scenarios and the different benign application traffic that each dataset contain. Therefore, the influence and importance of each network feature vary in each dataset. However, there are some common findings across datasets, such as the RF classifier being heavily influenced by forward directional-based features in the CICFlowMeter feature set.

\section{Conclusion}
ML-based NIDSs have achieved outstanding attack detection performance in the research community. However, the number of operational deployments has been very small. The limited evaluation of a common feature set across multiple datasets and the missing explanation of classification results contribute to the failed translation of research into real-world deployments. In this paper, the proposed NetFlow-based feature set has been evaluated and compared with the feature set designed by the CICFlowMeter tool. The evaluation has been conducted over three datasets (CSE-CIC-IDS2018, ToN-IoT, and BoT-IoT) using two ML classifiers (RF and DFF). Two datasets, CIC-BoT-IoT and CIC-ToN-IoT, have been generated and published to conduct the experiments. This kind of reliable comparison demonstrates the importance and necessity of having common feature sets across NIDS datasets, such as evaluating the generalisability of ML models' performance across different network environments and attack scenarios. The classification results generated by the ML models indicate a constant superiority of the NetFlow feature across the three NIDS datasets. Where both the DFF and RF classifiers achieved a higher attack detection accuracy in a lower prediction time. In addition, the SHAP method is used to explain the prediction results of the ML models by measuring the feature importance. The key features influencing the predictions of the models have been identified for each dataset.

\printbibliography

\end{document}